\documentclass[11pt,a4paper]{article}
\usepackage{amsmath}
\usepackage{amssymb}
\usepackage{CJK}
\usepackage{epsfig}
\usepackage{graphicx}
\usepackage[dvipdfm]{hyperref}
\usepackage[numbers,sort&compress,super]{natbib}

\makeatletter
\long\def\@makecaption#1#2{%
\vskip\abovecaptionskip
\sbox\@tempboxa{{\bfseries#1.}\hskip 1em#2}%
\ifdim \wd\@tempboxa >\hsize
   {\bfseries#1.}\hskip 1em#2\par
\else
  \global \@minipagefalse
  \hb@xt@\hsize{\hfil\box\@tempboxa\hfil}%
\fi \vskip\belowcaptionskip} \makeatother

\date{}

\begin{document}

\author{\small {Wei-Qiang Yang$^1$, Ya-Bo Wu$^1$\footnote{\href{mailto:ybwu61@163.com}
{Corresponding author: ybwu61@163.com}} , Li-Min Song$^2$, Yang-Yang
Su$^1$, Jian Li$^1$, Dan-Dan Zhang$^1$, Xiao-Gang Wang$^1$}}
\title{ \bf {Reconstruction of new holographic scalar field models of dark energy in Brans-Dicke universe}}
 \maketitle
 { \small { \centerline{\small{$^1$ Department of Physics,
Liaoning Normal University, Dalian 116029, P.R.China}} \small {
\centerline{\small{$^2$ College of Information Science and
Technology, Dalian Maritime University, Dalian 116026, P.R.China}}
\small{\centerline{\small{ woshiyangweiqiang@163.com ;
ybwu61@163.com}}} \vspace{1cm} \noindent{\small Motivated by the
work [K. Karami, J. Fehri, {{\it Phys. Lett. B}} {\bf 684}, 61
(2010)] and [A. Sheykhi, {{\it Phys. Lett. B}} {\bf 681}, 205
(2009)], we generalize their work to the new holographic dark energy
model with $\rho_D=\frac{3\phi^2}{4\omega}(\mu H^2+\nu\dot{H})$ in
the framework of Brans-Dicke cosmology. Concretely, we study the
correspondence between the quintessence, tachyon, K-essence, dilaton
scalar field and Chaplygin gas model with the new holographic dark
energy model in the non-flat Brans-Dicke universe. Furthermore, we
reconstruct the potentials and dynamics for these models. By
analysis we can show that for new holographic quintessence and
Chaplygin gas models, if the related parameters to the potentials
satisfy some constraints, the accelerated expansion can be achieved
in Brans-Dicke cosmology. Especially the counterparts of fields and
potentials in general relativity can describe accelerated expansion
of the universe. It is worth stressing that not only can we give
some new results in the framework of Brans-Dicke cosmology, but also
the previous results of the new holographic dark energy in Einstein
gravity can be included as special cases given by us.


\noindent{\sl PACS~: 95.36.+x, 98.80.-k\\}

According to the recent observations of type Ia supernovae\cite{1},
we learn that the universe is undergoing an accelerated expansion,
also along with the observations of CMBR anisotropy spectrum\cite{2}
and large scale structure (LSS)\cite{3}. The fact implies that there
must be some unknown component in universe which can drive the
accelerated expansion of the universe. This component is named as
dark energy (DE), and holds a large negative pressure. After this,
lots of candidates to dark energy have been suggested: (i) the
cosmological constant\cite{4}; (ii) scalar fields such as
quintessence\cite{5}, phantom\cite{6}, K-essence \cite{7}, tachyon
field \cite{8}, dilatonic ghost condensate \cite{9}, and so forth;
(iii) The interacting DE models including Chaplygin gas \cite{10},
generalized Chaplygin gas \cite{11,12,13}, braneworld models
\cite{14} and agegraphic DE models \cite{15,16}, etc.. Besides,
holographic dark energy (HDE) model \cite{17} is proposed on the
base of the holographic ideas. According to the holographic
principle, the number of degrees of freedom of a physical system
scales with the area of its boundary. In this context, Cohen et
al.\cite{18} suggested that one uses the event horizon as the
cosmological horizon, and the total energy in a region of size L
should not exceed the mass of a black hole of the same size. The
holographic DE not only gives the observational value of DE in the
universe, but also can drive the universe to an accelerated
expansion phase. In that case, however, an obvious drawback
concerning causality appears in this propose. This motivated Granda
and Oliveros \cite{19} to propose a new infrared cut-off for HDE,
which is about the square of the Hubble parameter $H^2$ and the time
derivative of the Hubble parameter $\dot{H}$, this model is called
as new HDE (NHDE) model in this paper. The NHDE model can avoid the
problem of causality which appears using the event horizon area as
the cut-off. By use of the new infrared cut-off for HDE, Granda and
Oliveros \cite{20} study the correspondence between the
quintessence, tachyon, K-essence and dilaton energy density with
NHDE model in the flat FRW universe. Subsequently, Karami and Fehri
\cite{21} generalize their work to the non-flat case.

On the other hand, it is quite possible that gravity is not given by
the Einstein action, at least at sufficiently high energies. In
string theory, gravity becomes scalar-tensor in nature. The low
energy limit of string theory leads to the Einstein gravity, coupled
non-minimally to a scalar field \cite{22}. Although the pioneering
study on scalar-tensor theories was done by Brans and Dicke several
decades ago who applied Mach's principle into gravity \cite{23}, it
has got a new impetus as it arises naturally as the low energy limit
of many theories of quantum gravity (for example, superstring
theory, et al.), and scalar-tensor theories of gravity have been
widely applied in cosmology \cite{24,25,26}. For HDE as a dynamical
model, we need a dynamical frame to accommodate it instead of
general relativity. Recently, the studies on the HDE model in the
framework of Brans-Dicke theory have been carried out
\cite{27,28,29,30,31}, the purpose of which is to construct a
cosmological model of late acceleration based on the Brans-Dicke
theory of gravity.

In this paper, we will try to generalize the previous work in
Refs.[21] and [27] to the NHDE model with
$\rho_D=\frac{3\phi^2}{4\omega}(\mu H^2+\nu\dot{H})$ in the
framework of Brans-Dicke cosmology. Hence, the evolution of equation
of state (EoS) for the NHDE model will be discussed in non-flat
Brans-Dicke universe. Moreover, we will study the correspondence
between the quintessence, tachyon, K-essence, dilaton and Chaplygin
gas model with NHDE model in the non-flat Brans-Dicke universe.
Furthermore, we will reconstruct the potentials and dynamics for
these models. By analysis we can show that for new holographic
quintessence and Chaplygin gas models, if the related parameters to
the potentials satisfy some constraints, the accelerated expansion
of universe can be achieved in Brans-Dicke cosmology.

We start from the action of Brans-Dicke theory, in the canonical
form it can be written \cite{32}
\begin{equation} \label{1}
S=\int d^4x\sqrt{g}(-\frac{1}{8\omega} \phi^2
R+\frac{1}{2}g^{\mu\nu}
\partial_\mu \partial_\nu \phi+L_M),
\end{equation}
where $R$ is the scalar curvature and $\phi$ is the Brans-Dicke
scalar field. The non-minimal coupling term $\phi^2 R$ replaces with
the Einstein-Hilbert term $R/G$ in such a way that $G^{-1}=2\pi
\phi^2/\omega$, where $G$ is the gravitational constant. The signs
of the non-minimal coupling term and the kinetic energy term are
properly adopted to $(+ - - -)$ metric signature. Here, we consider
the non-flat Friedmann-Robertson-Walker (FRW) universe which is
described by the line element
\begin{equation} \label{2}
ds^2=dt^2-a^2(t)(\frac{dr^2}{1-kr^2}+r^2d\Omega ^{2}),
\end{equation}
where $a(t)$ is the scale factor, and $k$ is the curvature parameter
with $k=-1,0,1$ corresponding to open, flat, and closed universes,
respectively. Varying action (\ref{1}) with respect to metric
(\ref{2}), one can obtain the following field equations
\begin{equation}\label{3}
\frac{3}{4\omega}\phi^2(H^2+\frac{k}{a^2})-\frac{1}{2}\dot{\phi}^2+\frac{3}{2\omega}H\dot{\phi}\phi
=\rho_D,
\end{equation}
\begin{equation}\label{4}
\frac{-1}{4\omega}\phi^2(2\frac{\ddot{a}}{a}+H^2+\frac{k}{a^2})-\frac{1}{\omega}H\dot{\phi}\phi-
\frac{1}{2\omega}\ddot{\phi}\phi-\frac{1}{2}(1+\frac{1}{\omega})\dot{\phi}^2=p_D,
\end{equation}
\begin{equation}\label{5}
\ddot{\phi}+3H\dot{\phi}-\frac{3}{2\omega}(\frac{\ddot{a}}{a}+H^2+\frac{k}{a^2})\phi=0,
\end{equation}
where the dot is the derivative with respect to time and
$H=\dot{a}/a$ is the Hubble parameter. Here $\rho_D$ and $p_D$ are,
respectively, the density and pressure of DE. And we neglect the
contributions from matter and radiation in the universe, that is,
$\rho_{total}=\rho_D$. In addition, we shall assume that Brans-Dicke
field can be described as a power law of the scale factor,
$\phi\propto a^\alpha$. Taking the derivative with respect to time,
one can get
\begin{equation}\label{6}
\dot{\phi}=\alpha H\phi,
\end{equation}
\begin{equation}\label{7}
\ddot{\phi}=\alpha^2H^2\phi+\alpha\phi\dot{H}.
\end{equation}

According to Ref.[16], the HDE density with the new infrared cut-off
is given by
\begin{equation}\label{8}
\rho_D=\frac{3}{8\pi G}(\mu H^2+\nu\dot{H}),
\end{equation}
this is just the NHDE model, where $H=\dot{a}/a$ is the Hubble
parameter, $\mu$ and $\nu$ are constants which must satisfy the
restrictions imposed by the current observational data.

In the framework of Brans-Dicke theory, using $\phi^2=\omega/2\pi
G_{eff}$ , one can obtain
\begin{equation}\label{9}
\rho_D=\frac{3\phi^2}{4\omega}(\mu H^2+\nu\dot{H}).
\end{equation}

Thus, Eq.({\ref{3}}) becomes
\begin{equation}\label{10}
\frac{3}{4\omega}\phi^2(H^2+\frac{k}{a^2})-\frac{1}{2}\dot{\phi}^2+\frac{3}{2\omega}H\dot{\phi}\phi
=\frac{3\phi^2}{4\omega}(\mu H^2+\nu\dot{H}).
\end{equation}

Furthermore, we obtain
\begin{equation}\label{11}
\frac{dH^2}{dx}+\frac{2s}{3\nu}H^2=\frac{2k}{\nu}e^{-2x},
\end{equation}
where $x=lna$ and $s=2\omega\alpha^2-6\alpha+3\mu-3$. Integrating
the above equation with respect to $x$ yields
\begin{equation}\label{12}
H^2=\frac{3k}{s-3\nu}e^{-2x}+Ce^{-2sx/3\nu},
\end{equation}
where $C$ is an integration constant and $s\neq3\nu$. For the
special case of $s=3\nu$, from Eq.(\ref{11}), we have
$H^2=\frac{2k}{\nu}xe^{-2x}+Ce^{-2x}$. But, here we only make the
discussion of $s\neq3\nu$. For the flat case $k=0$ and using
$\dot{x}=H$, from Eq.(\ref{11}), one can reduce to
\begin{equation}\label{13}
H^2=-\frac{3\nu}{s}\dot{H},
\end{equation}
where the dot denotes the time derivative with respect to the cosmic
time $t$. Furthermore, one has
\begin{equation}\label{14}
H=\frac{3\nu}{s}\frac{1}{t}.
\end{equation}

It is easy to see from Eq.(\ref{14}) that when $\alpha=0$, Hubble
parameter $H=\frac{\nu}{\mu-1}\frac{1}{t}$, which  exactly  reduces
to the case in Ref.[20].

In addition, according to the conservation equation
\begin{equation}\label{15}
\dot{\rho}_D+3H(1+w_D)\rho_D=0,
\end{equation}
and $w_D=p_D/\rho_D$, EoS can be expressed as
\begin{equation}\label{16}
w_D=-1-\frac{2\alpha\mu
H^3+2(\alpha\nu+\mu)H\dot{H}+\nu\ddot{H}}{3H(\mu H^2+\nu\dot{H})},
\end{equation}
which is just the evolution of EoS for the NHDE model in non-flat
Brans-Dicke universe. When $\alpha=0$, it is easy to see that the
Brans-Dicke scalar field becomes trivial, and Eq.(\ref{16}) can
reduce to the case of NHDE model in general relativity \cite{20,21},
i.e.,
\begin{equation}\label{17}
w_D=-1-\frac{2\mu H\dot{H}+\nu\ddot{H}}{3H(\mu H^2+\nu\dot{H})}.
\end{equation}

It follows that the results in Refs.[20,21] are included as the
special cases of $\alpha=0$ and $k=0$ given by us in this paper.

Furthermore, from Eqs.(\ref{12}) and (\ref{16}) we obtain
\begin{equation}\label{18}
w_D=\frac{27e^{2sx/3\nu}k\nu(1+2\alpha)(\mu-\nu)+X(2s-9\nu-6\alpha\nu)}
{9\nu[X+9e^{2sx/3\nu}k(\nu-\mu)]},
\end{equation}
where $X=Ce^{2x}(s-3\mu)(s-3\nu)$ (Here, $s\neq3\nu$ and
$\mu\neq\nu$). Eq.(\ref{18}) shows that EoS parameter is
time-dependent, and can cross the phantom divide $w_D=-1$
\cite{33,34,35}. Of course, the current observations from the
seven-year WMAP data and the analysis of SN1a and CMB data \cite{36}
favor the present values $w_0$ of EoS bigger than $-1$.

The evolutionary trajectories of EoS in Eq.(\ref{18}) are plotted in
Fig.{\ref{F1}} (Here we take $k=1$, $\omega=-1000$ and
$\alpha=-1/1000$). From Fig.{\ref{F1}}, it is easy to see that the
EoS of NHDE model can cross the phantom divide $w=-1$. Furthermore,
we also give the present values of EoS $w_0$ in Tab.{\ref{T1}}. The
values of $w_{0}$ are basically in the ranges of $w_0=-1.10\pm0.14$,
which are supported by the seven-year WMAP data and the analysis of
SN1a and CMB data \cite{36}.

\begin{figure}[!htb] \vspace{-0.6cm} \hspace{-0.6cm}
\centering
\includegraphics[width=170pt,height=136pt]{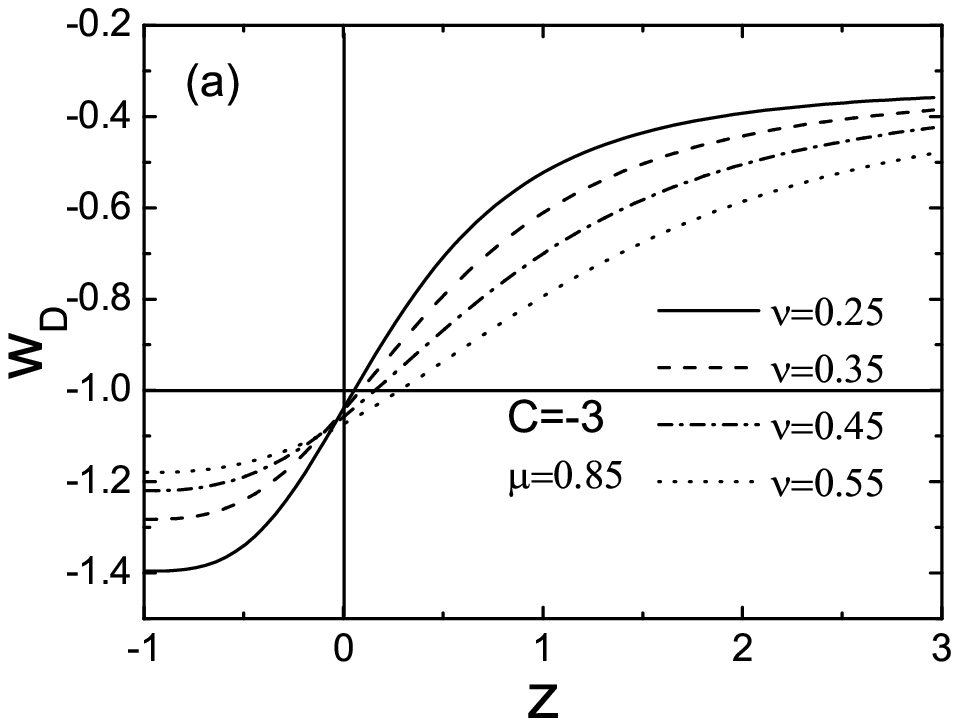} \hspace{-1.2cm}
\includegraphics[width=170pt,height=136pt]{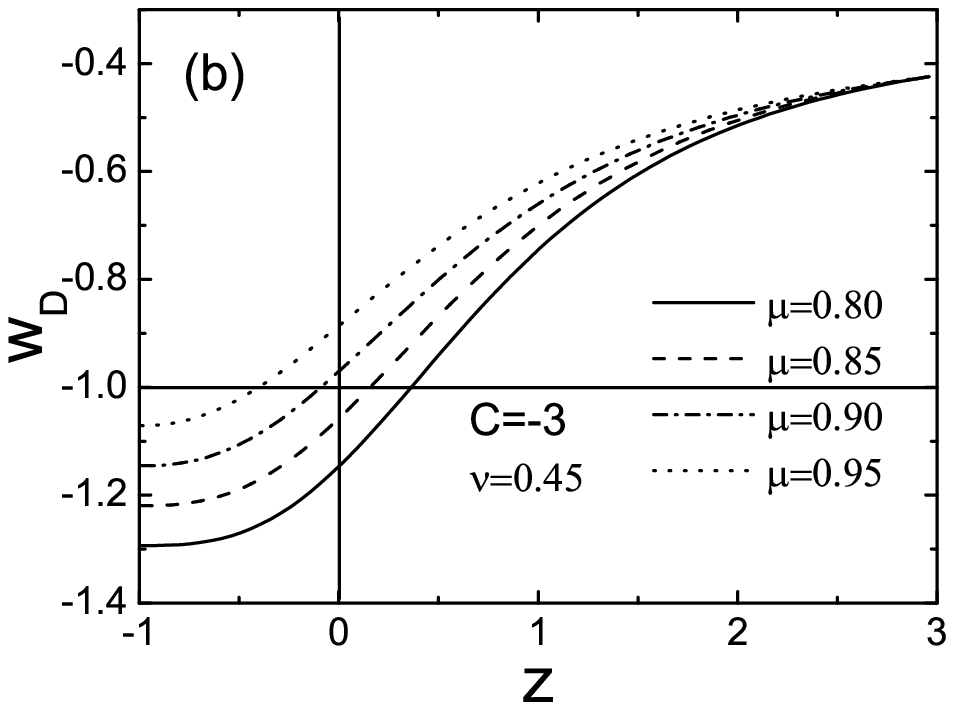} \hspace{-1.2cm}
\includegraphics[width=170pt,height=136pt]{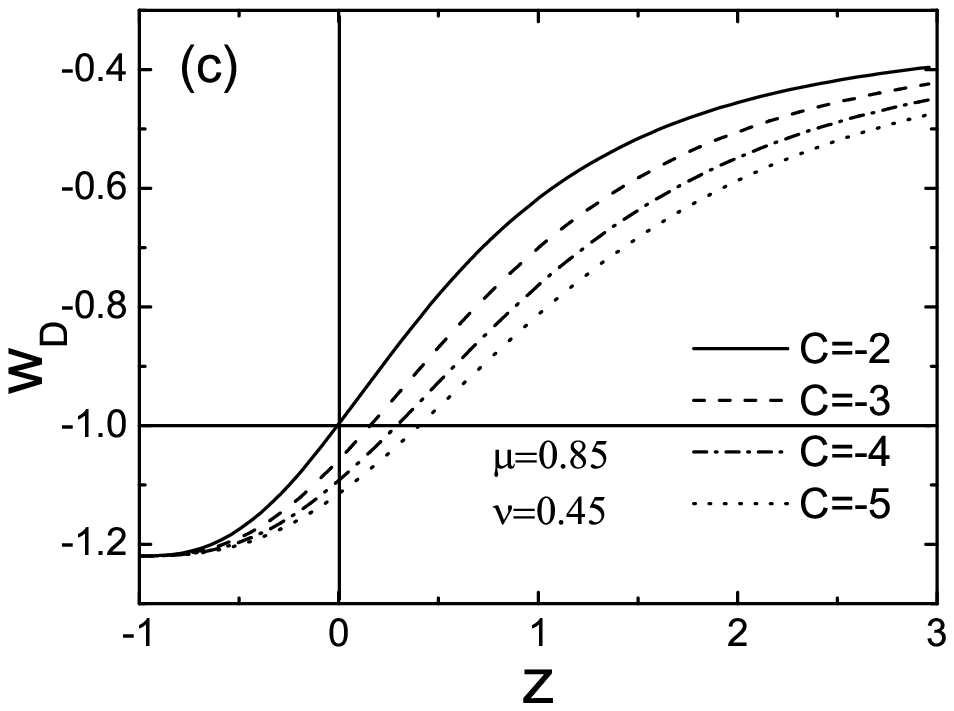} \vspace{-0.6cm}
\caption{\small{The evolutionary trajectories of EoS in NHDE model.
(a) for fixed constants $C$ and $\mu$ but different coupling
constants $\nu$; (b) for fixed constants $C$ and $\nu$ but different
constants $\mu$; (c) for fixed constants $\mu$ and $\nu$ but
different constants $C$. }} \label{F1}
\end{figure}

\begin{table}[!htb]
\centering \hspace{0.1cm}\small{\begin{tabular}{|c|c|c|}
\hline $w_0(C=-3,\mu=0.85)$ & $w_0(C=-3,\nu=0.45)$ & $w_0(\mu=0.85,\nu=0.45)$ \\
\hline $-1.04030(\nu=0.25)$ & $-1.14700(\mu=0.80)$ & $-1.99726(C=-2)$ \\
\hline $-1.04433(\nu=0.35)$ & $-1.05785(\mu=0.85)$ & $-1.05785(C=-3)$ \\
\hline $-1.05785(\nu=0.45)$ & $-0.97081(\mu=0.90)$ & $-1.09249(C=-4)$ \\
\hline $-1.07338(\nu=0.55)$ & $-0.88619(\mu=0.95)$ & $-1.11491(C=-5)$ \\
\hline
\end{tabular}}
\caption{\small{The present values $w_0$ of EoS in Fig.
{\ref{F1}}.}}  \label{T1}
\end{table}

\newpage

We know that Brans-Dicke cosmology becomes standard cosmology when
$\omega\rightarrow \infty$, in this case $\alpha\rightarrow 0$,
according to this result, we can obtain
\begin{equation}\label{19}
w_D=-\frac{1}{3}\left(\frac{k(\frac{\mu-\nu}{\mu-\nu-1})e^{-2x}+
C(\frac{3\nu-2\mu+2}{\nu})e^{-\frac{2}{\nu}(\mu-1)x}}{k(\frac{\mu-\nu}{\mu-\nu-1})e^{-2x}
+Ce^{-\frac{2}{\nu}(\mu-1)x})}\right)
\end{equation}
which is just the same as Eq.(10) in Ref.[21].

If taking $k=0$ for the flat universe, we obtain
\begin{equation}\label{20}
w_D=-1-\frac{2\alpha}{3}+\frac{2s}{9\nu}.
\end{equation}

In addition, by use of $w_D<-1/3$ and the present values $H_0$,
$t_0$, $a_0$, we can obtain the constraints of the parameters (such
as $\mu$, $\nu$, $\omega$ and $\alpha$ etc.). For the non-flat case
($k\neq0$), in Eq.(\ref{12}), when taking $H=H_0$, $a=a_0=1$ (i.e,
$x_0=lna_0=0$), the integration constant $C$ can be expressed as
$C=H^2_0-3k/(s-3\nu)$. Thus, by means of EoS parameter $w_D<-1/3$ in
Eq.(\ref{18}), we get the following constraint
$\frac{3k(-6-3\nu-12\alpha-6\alpha\nu+4\omega\alpha^2)
-H^2_0(2\omega\alpha^2-6\alpha-3)(-6+6\mu-9\nu-12\alpha-6\alpha\nu+4\omega\alpha^2)}
{27k\nu+9H^2_0(3+6\alpha-2\omega\alpha^2)}<-1/3$. Specially, for the
flat case of $k=0$, from Eqs.(\ref{14}) and (\ref{20}), we can
obtain $H_0t_0=3\nu/s$, and
$w_D=-1-\frac{2\alpha}{3}+\frac{2}{3H_0t_0}<-1/3$, which leads to
$\alpha>\frac{1}{H_0t_0}-1$. Furthermore, according to
$H_0t_0=\frac{1}{3\sqrt{\Omega_D}}ln\left[\frac{1+\sqrt{\Omega_D}}{1-\sqrt{\Omega_D}}\right]$
\cite{37,38}, we find when $\Omega_D\rightarrow1$,
$H_0t_0\rightarrow\infty$. Hence, we can conclude that for the flat
case of $k=0$, the constraint on $\alpha$ is $\alpha>-1$.

Contrary to the non-flat case, the EoS for the flat case is
constant. Furthermore, when $\alpha=0$, Eq.(\ref{20}) reduces to the
case of NHDE model in general relativity \cite{20}
\begin{equation}\label{21}
w_D=-1+\frac{2(\mu-1)}{3\nu}.
\end{equation}

In order to obtain accelerated expansion, the constants $\mu$ and
$\nu$ must satisfy the restrictions: for the quintessence-like phase
with $-1<w_D<-1/3$, we obtain $\mu>1$ and $\nu>\mu-1$, or, $\mu<1$
and $\nu<\mu-1$. For $\mu<1$ and $\nu>0$, or, $\mu>1$ and $\nu<0$,
it describes a phantom-like phase with $w_D<-1$.

Below, we will study the correspondence between NHDE model with
quintessence, tachyon, K-essence and dilaton scalar field models as
well as Chaplygin gas model in the non-flat Brans-Dicke universe.
Also, we will reconstruct the potentials and dynamics for these
scalar field models. Specially, when taking $k=0$, we can give the
related results of scalar fields and potentials for the NHDE model
in the flat Brans-Dicke universe. Furthermore, we can obtain the
fields and potentials in general relativity, which describe
accelerated expansion of the universe. To establish this
correspondence, we compare energy density of the NHDE model given by
Eq.(\ref{9}) with the corresponding energy density of scalar field
model, and also equate the EoS for these scalar models with the EoS
given by Eq.(\ref{18}). Here, we take scalar field as $\varphi$ in
order to differ from the parameter $\phi$ in Brans-Dicke theory.

1. New holographic quintessence model

The energy density and pressure of the quintessence scalar field
$\varphi$ are as follows \cite{5}
\begin{equation} \label{22}
\rho_Q=\frac{1}{2}\dot{\varphi}^2+V(\varphi),
\end{equation}
\begin{equation} \label{23}
p_Q=\frac{1}{2}\dot{\varphi}^2-V(\varphi).
\end{equation}
The EoS for the quintessence scalar field is given by
\begin{equation} \label{24}
w_Q=\frac{p_Q}{\rho_Q}=\frac{\dot{\varphi}^2-2V(\varphi)}{\dot{\varphi}^2+2V(\varphi)}.
\end{equation}

Therefore, we have
\begin{equation} \label{25}
w_D=\frac{27e^{2sx/3\nu}k\nu(1+2\alpha)(\mu-\nu)+X(2s-9\nu-6\alpha\nu)}
{9\nu[X+9e^{2sx/3\nu}k(\nu-\mu)]}=
\frac{\dot{\varphi}^2-2V(\varphi)}{\dot{\varphi}^2+2V(\varphi)},
\end{equation}
and
\begin{equation} \label{26}
\rho_D=\frac{3\phi^2}{4\omega}(\alpha
H^2+\beta\dot{H})=\frac{1}{2}\dot{\varphi}^2+V(\varphi).
\end{equation}

Furthermore, the kinetic energy term and the quintessence potential
energy are
\begin{equation} \label{27}
\dot{\varphi}^2=\frac{e^{2(\alpha-s/3\nu-1)x}[27e^{2sx/3\nu}k\nu(1-\alpha)(\mu-\nu)+
X(3\alpha\nu-s)]}{18\omega\nu(s-3\nu)},
\end{equation}
\begin{equation} \label{28}
V(\varphi)=\frac{e^{2(\alpha-s/3\nu-1)x}[27e^{2sx/3\nu}k\nu(2+\alpha)(\mu-\nu)
+X(s-9\nu-3\alpha\nu)]}{36\omega\nu(s-3\nu)}.
\end{equation}

Using $\dot{\varphi}=H\varphi'$, where prime denotes the derivative
with respect to $x$, we obtain
\begin{equation} \label{29}
\varphi'=\frac{1}{H}\sqrt{\frac{e^{2(\alpha-s/3\nu-1)x}[27e^{2sx/3\nu}k\nu(1-\alpha)(\mu-\nu)+
X(3\alpha\nu-s)]}{18\omega\nu(s-3\nu)}},
\end{equation}
where $H$ is given by Eq. (\ref{12}). Consequently, after
integration with respect to $x$ we can obtain the evolutionary form
of the quintessence scalar field as
\begin{equation} \label{30}
\varphi(a)-\varphi(0)=\int^{lna}_{0}\frac{1}{H}\sqrt{\frac{e^{2(\alpha-s/3\nu-1)x}
[27e^{2sx/3\nu}k\nu(1-\alpha)(\mu-\nu)+X(3\alpha\nu-s)]}{18\omega\nu(s-3\nu)}}dx,
\end{equation}
where we take $a_0=1$ for the present time.

For the flat case of $k=0$, assuming $\varphi (0)=0$ for the present
$t_0=0$, then Eqs.(\ref{28}) and (\ref{30}) reduce to
\begin{equation} \label{31}
\varphi(t)=\sqrt{\frac{(s-3\mu)(3\alpha\nu-s)}{18\omega\nu}}\frac{t^{3\alpha\nu/s}}{\alpha}~~~~~~~~~(\alpha\neq0),
\end{equation}
\begin{equation} \label{32}
V(\varphi)=\frac{\nu(s-3\mu)(s-9\nu-3\alpha\nu)}{4\omega
s^2}\left[\sqrt{\frac{18\omega\nu}{(s-3\mu)(3\alpha\nu-s)}}\alpha\varphi\right]
^{2(1-s/3\alpha\nu)}.
\end{equation}

It is easy to see that when taking $3\alpha\nu/s=-1$, Eq.(\ref{32})
can reduce to $V=V_1\varphi^4$ ($V_1$ is a constant) which describes
the accelerated expansion of universe \cite{39}. By using the
decelerated parameter $q\equiv-a\ddot{a}/\dot{a}^2<0$ and
$a=t^{-2/\alpha}$, we have $-2<\alpha<0$ or $\alpha>0$. Furthermore,
$w_D=-1-2\alpha/3+2s/9\nu<-1/3$ can give $\alpha>-1/2$, so we obtain
$-1/2<\alpha<0$ or $\alpha>0$. In addition, considering that
$\frac{18\omega\nu}{(s-3\mu)(3\alpha\nu-s)}>0$ in Eqs.(\ref{31}) and
(\ref{32}), thus we have $\omega<-3/2$ for $-1/2<\alpha<0$ and
exclude the case of $\alpha>0$. Therefore we believe that the
potential (\ref{32}) for the new holographic quintessence model,
which can reduce to $V=V_1\varphi^4$,  can describe the accelerated
expansion of universe if the parameters satisfy the constraints
$3\alpha\nu/s=-1$, $-1/2<\alpha<0$ and $\omega<-3/2$.

Specially, when taking $k=0$ and $\alpha=0$ in Eqs.(\ref{28}) and
(\ref{29}), the scalar field and potential can reduce to the case of
NHDE model in general relativity \cite{20}
\begin{equation} \label{33}
\varphi(t)=\sqrt{\frac{2\nu}{\mu-1}}M_plnt,
\end{equation}
\begin{equation} \label{34}
V(\varphi)=\frac{3\nu-\mu+1}{(\mu-1)^2}M^2_p
exp\left(-\sqrt{\frac{2(\mu-1)}{\nu}}\frac{\varphi}{M_p}\right).
\end{equation}
According to Ref.[40], this potential can describe an accelerated
expansion provided that $\nu/(\mu-1)>1$, and also has cosmological
scaling solutions \cite{41}.

2. New holographic tachyon model

The energy density and pressure for the tachyon field are as follows
\cite{8}
\begin{equation} \label{35}
\rho_T=\frac{V(\varphi)}{\sqrt{1-\dot{\varphi}^2}},
\end{equation}
\begin{equation} \label{36}
p_T=-V(\varphi)\sqrt{1-\dot{\varphi}^2},
\end{equation}
where $V(\varphi)$ is tachyon potential. The EoS for the tachyon
scalar field is obtained as
\begin{equation} \label{37}
w_T=\frac{p_T}{\rho_T}=\dot{\varphi}^2-1.
\end{equation}

By means of $w_D=w_T$ and $\rho_D=\rho_T$, we can obtain the kinetic
energy term and the tachyon potential energy in the new holographic
tachyon model as follows
\begin{equation} \label{38}
\dot{\varphi}^2=\frac{54e^{2sx/3\nu}k\nu(\alpha-1)(\mu-\nu)+2X(s-3\alpha\nu)}
{9\nu[X+9e^{2sx/3\nu}k(\nu-\mu)]},
\end{equation}
\begin{equation} \label{39}
V(\varphi)=\frac{e^{2(\alpha-s/3\nu-1)x}[9e^{2sx/3\nu}k(\mu-\nu)-X]Y}
{12\omega(s-3\nu)},
\end{equation}
where
$Y=\sqrt{\frac{27e^{2sx/3\nu}k\nu(1+2\alpha)(\nu-\mu)-X(2s-9\nu-6\alpha\nu)}
{\nu[X+9e^{2sx/3\nu}k(\nu-\mu)]}}$.

Furthermore, the evolutionary form of the tachyon scalar field can
be obtained
\begin{equation} \label{40}
\varphi(a)-\varphi(0)=\int^{lna}_{0}\frac{1}{H}\sqrt{\frac{54e^{2sx/3\nu}k\nu(\alpha-1)(\mu-\nu)+
2X(s-3\alpha\nu)} {9\nu[9e^{2sx/3\nu}k(\nu-\mu)+X]}}dx.
\end{equation}

For the flat case $k=0$, using the initial condition $\varphi
(0)=0$, we have
\begin{equation} \label{41}
\varphi(t)=\sqrt{\frac{2(s-3\alpha\nu)}{9\nu}}t,
\end{equation}
\begin{equation} \label{42}
V(\varphi)=\frac{3(3\mu-s)\nu^2}{4\omega
s^2}\sqrt{\frac{9\nu+6\alpha\nu-2s}{\nu}}\left[\sqrt{\frac{9\nu}{2(s-3\alpha\nu)}}\varphi\right]^
{2(3\alpha\nu/s-1)}.
\end{equation}
Furthermore, when taking $\alpha=0$, Eqs.(\ref{41}) and (\ref{42})
can reduce to the counterparts in general relativity \cite{20}
\begin{equation} \label{43}
\varphi(t)=\sqrt{\frac{2(\mu-1)}{3\nu}}t,
\end{equation}
\begin{equation} \label{44}
V(\varphi)=2M^2_p\sqrt{1-\frac{2(\mu-1)}{3\nu}}\left(\frac{\nu}{\mu-1}\right)\frac{1}{\varphi^2}.
\end{equation}
According to Refs.[20] and [40], the inverse square potential in
Eq.(\ref{44}) can describe the accelerated expansion of universe
when the parameters satisfy the condition
$-1/9(1+\sqrt{10})<(\mu-1)/\nu<1/9(-1+\sqrt{10})$, which gives the
only viable late-time attractor solution.

3. New holographic K-essence model

The energy density and pressure for the K-essence DE model are given
by \cite{7}
\begin{equation} \label{45}
\rho(\varphi,\chi)=f(\varphi)(-\chi+3\chi^2),
\end{equation}
\begin{equation} \label{46}
p(\varphi,\chi)=f(\varphi)(-\chi+\chi^2).
\end{equation}

The EoS is obtained as
\begin{equation} \label{47}
w_{KE}=\frac{p(\varphi,\chi)}{\rho(\varphi,\chi)}=\frac{\chi-1}{3\chi-1}.
\end{equation}

Equating Eq.(\ref{47}) with the EoS of NHDE, $w_{KE}=w_D$, one has
\begin{equation} \label{48}
\chi=\frac{27e^{2sx/3\nu}k\nu(\alpha+2)(\mu-\nu)+X(s-9\nu-3\alpha\nu)}
{3[27e^{2sx/3\nu}k\nu(\alpha+1)(\mu-\nu)+X(s-6\nu-3\alpha\nu)]}.
\end{equation}

By use of Eq.(\ref{48}), $\dot{\varphi}^2=2\chi$, and
$\dot{\varphi}=\varphi'H$, we obtain the evolutionary form of the
K-essence scalar field as
\begin{equation} \label{49}
\varphi(a)-\varphi(0)=
\int^{lna}_{0}\frac{1}{H}\sqrt{\frac{2[27e^{2sx/3\nu}k\nu(\alpha+2)(\mu-\nu)+X(s-9\nu-3\alpha\nu)]}
{3[27e^{2sx/3\nu}k\nu(\alpha+1)(\mu-\nu)+X(s-6\nu-3\alpha\nu)]}}dx.
\end{equation}

Furthermore, using $\rho_D=\rho(\varphi,\chi)$, Eqs. (\ref{3}) and
(\ref{12}), the expression of $f(\varphi)$ can be given as follows
\begin{equation} \label{50}
f(\varphi)=\frac{e^{2(\alpha-s/3\nu-1)x}[9e^{2sx/3\nu}k(\mu-\nu)-X]}
{4\omega(s-3\nu)\chi(3\chi-1)}.
\end{equation}

For the flat case of $k = 0$, assuming $\varphi(0)=0$, then Eqs.
(\ref{48}), (\ref{49}) and (\ref{50}) become into
\begin{equation} \label{51}
\chi=\frac{s-9\nu-3\alpha\nu}{3(s-6\nu-3\alpha\nu)},
\end{equation}
\begin{equation} \label{52}
\varphi(t)=\sqrt{\frac{2(s-9\nu-3\alpha\nu)}{3(s-6\nu-3\alpha\nu)}}t,
\end{equation}
\begin{equation} \label{53}
f(\varphi)=\frac{9\nu(s-3\mu)(s-6\nu-3\alpha\nu)^2}{4\omega
s^2(s-9\nu-3\alpha\nu)}\left[\sqrt{\frac{3(s-6\nu-3\alpha\nu)}{2(s-9\nu-3\alpha\nu)}}\varphi\right]
^{2(3\alpha\nu/s-1)},
\end{equation}
Thus, when taking $\alpha=0$, Eqs.(\ref{52}) and (\ref{53}) can
reduce to
$\varphi(t)=\sqrt{\frac{2}{3}\left(\frac{3\nu-\mu+1}{2\nu-\mu+1}\right)}t$
and
$f(\varphi)=6M^2_p\nu\left[\frac{2\nu-\mu+1}{(\mu-1)^2}\right]\frac{1}{\varphi^2}$
\cite{20}. Hence, the potential of new holographic K-essence
corresponds to the power-law expansion like new holographic tachyon
model.

4. New holographic dilaton model

The energy density and pressure of the dilaton DE model are given by
\cite{9}
\begin{equation} \label{54}
\rho_{DI}=\chi-3c'e^{\lambda\varphi}\chi^2,
\end{equation}
\begin{equation} \label{55}
p_{DI}=\chi-c'e^{\lambda\varphi}\chi^2,
\end{equation}
where $c'$ and $\lambda$ are positive constants and
$\chi=\dot{\varphi}^2/2$. The EoS for the dilaton scalar field is
given by
\begin{equation} \label{56}
w_{DI}=\frac{p_{DI}}{\rho_{DI}}=\frac{\chi-c'e^{\lambda\varphi}\chi^2}{\chi-3c'e^{\lambda\varphi}\chi^2}.
\end{equation}

By means of $\omega_{DI}=\omega_D$, we find the following solution
\begin{equation} \label{57}
c'e^{\lambda\varphi}\chi=\frac{27e^{2sx/3\nu}k\nu(\alpha+2)(\mu-\nu)+X(s-9\nu-3\alpha\nu)}
{3[27e^{2sx/3\nu}k\nu(\alpha+1)(\mu-\nu)+X(s-6\nu-3\alpha\nu)]}.
\end{equation}

Moreover, using $\chi=\dot{\varphi}^2/2$ and
$\dot{\varphi}=\varphi'H$, we have
\begin{equation} \label{58}
e^{\frac{\lambda\varphi(a)}{2}}=e^{\frac{\lambda\varphi(0)}{2}}+\frac{\lambda}{\sqrt{6c'}}\int^{lna}_{0}
\frac{1}{H}\sqrt{\frac{27e^{2sx/3\nu}k\nu(\alpha+2)(\mu-\nu)+X(s-9\nu-3\alpha\nu)}
{27e^{2sx/3\nu}k\nu(\alpha+1)(\mu-\nu)+X(s-6\nu-3\alpha\nu)}}dx,
\end{equation}
where H is given by Eq. (\ref{12}). Therefore, the evolutionary form
of the dilaton scalar filed can be written as
\begin{equation} \label{59}
\varphi(a)=\frac{2}{\lambda}ln[e^{\frac{\lambda\varphi(0)}{2}}+\frac{\lambda}{\sqrt{6c'}}\int^{lna}_{0}
\frac{1}{H}\sqrt{\frac{27e^{2sx/3\nu}k\nu(\alpha+2)(\mu-\nu)+X(s-9\nu-3\alpha\nu)}
{27e^{2sx/3\nu}k\nu(\alpha+1)(\mu-\nu)+X(s-6\nu-3\alpha\nu)}}dx].
\end{equation}

For the flat case $k=0$, using $\dot{x}=H$, assuming
$\varphi(0)\rightarrow-\infty$ for the present time $t_0=0$, then
one can give
\begin{equation} \label{60}
\varphi(t)=\frac{2}{\lambda}ln[\sqrt{\frac{s-9\nu-3\alpha\nu}
{6c'(s-6\nu-3\alpha\nu)}}\lambda t].
\end{equation}

By using $\alpha=0$, Eq.(\ref{60}) becomes into
$\varphi(t)=\frac{2}{\lambda}ln[\frac{\lambda}{\sqrt{6c'}}\sqrt{\frac{3\nu-\mu+1}
{2\nu-\mu+1}} t]$ \cite{20}. According to Ref.[40],  we know that
for the dynamics system of this model, the condition of accelerated
expansion of universe is $-2 < (\mu-1)/\nu < 1$.

5. New holographic Chaplygin gas model

Kamenshchik et al.\cite{10} proposed a model of DE involving a fluid
known as a Chaplygin gas(CG). This fluid also leads to the
acceleration of the universe at late times, and its simplest form
has the following specific equation of state
\begin{equation} \label{61}
p_{CG}=-\frac{A}{\rho_{CG}},
\end{equation}
where $A$ is a positive constant. With the continuity equation, Eq.
(\ref{61}) can be integrated to give
\begin{equation} \label{62}
\rho_{CG}=\sqrt{A+Be^{-6x}},
\end{equation}
where $B$ is an integration constant and $x=lna$. We will establish
the correspondence between NHDE model and CG model, that is, the new
holographic CG model. To do this, we treat CG as an ordinary scalar
field $\varphi$. So we have
\begin{equation} \label{63}
\rho_{\varphi}=\dot{\varphi}^2/2+V(\varphi)=\sqrt{A+Be^{-6x}},
\end{equation}
\begin{equation} \label{64}
p_{\varphi}=\dot{\varphi}^2/2-V(\varphi)=\frac{-A}{\sqrt{A+Be^{-6x}}}.
\end{equation}

Hence, it is easy to obtain the kinetic energy terms and the scalar
potential for the CG as
\begin{equation} \label{65}
\dot{\varphi}^2=\frac{Be^{-6x}}{a^6\sqrt{A+Be^{-6x}}},
\end{equation}
\begin{equation} \label{66}
V(\varphi)=\frac{2A+Be^{-6x}}{a^6\sqrt{A+Be^{-6x}}}.
\end{equation}

By use of $\rho_{CG}=\rho_D$, one can obtain
\begin{equation} \label{67}
B=\frac{e^{-4x/3\nu}[M-16Ae^{(6+4s/3\nu)x}\omega^2(s-3\nu)^2]}{16\omega^2(s-3\nu)^2},
\end{equation}
where
$M=C^2e^{(6+4\alpha)x}(s-3\mu)^2(s-3\nu)^2+81e^{(2+4\alpha+4s/3\nu)x}k^2(\mu-\nu)^2
+18Ce^{2x(6+6\alpha+s/\nu)/3}k(s-3\mu)(s-3\nu)(\nu-\mu)$. And using
$p_{CG}=\rho_{CG}w_D$, we have
\begin{equation} \label{68}
B=Ae^{6x}\left[-1-\frac{9\nu X+81\nu
e^{2sx/3\nu}k(\nu-\mu)}{27e^{2sx/3\nu}k\nu(1+2\alpha)(\mu-\nu)+X(2s-9\nu-6\alpha\nu)}\right].
\end{equation}
So it is easy to give the expression of $A$ and $B$
\begin{equation} \label{69}
A=-\frac{e^{-(6+4s/3\nu)x}M[27e^{2sx/3\nu}k\nu(1+2\alpha)(\mu-\nu)+X(2s-9\nu-6\alpha\nu)]}
{144\omega^2\nu(s-3\nu)^2[X+9e^{2sx/3\nu}k(\nu-\mu)]},
\end{equation}
\begin{equation} \label{70}
B=\frac{e^{-4sx/3\nu}M[27e^{2sx/3\nu}k\nu(\alpha-1)(\mu-\nu)
+X(s-3\alpha\nu)]}{72\omega^2\nu(s-3\nu)^2[X+9e^{2sx/3\nu}k(\nu-\mu)]}.
\end{equation}

Thus we can obtain the kinetic energy terms and the scalar potential
for the Chaplygin gas as
\begin{equation} \label{71}
\dot{\varphi}^2=\frac{[27e^{2sx/3\nu}k\nu(\alpha-1)(\mu-\nu)
+X(s-3\alpha\nu)]N}{18\nu[X+9e^{2sx/3\nu}k(\nu-\mu)]},
\end{equation}
\begin{equation} \label{72}
V(\varphi)=\frac{[27e^{2sx/3\nu}k\nu(\alpha+2)(\nu-\mu)
-X(s-9\nu-3\alpha\nu)]N}{36\nu[X+9e^{2sx/3\nu}k(\nu-\mu)]},
\end{equation}
where $N=\sqrt{\frac{e^{-(6+4s/3\nu)x}M}{\omega^2(s-3\nu)^2}}$.
Using $\dot{\varphi}=H\varphi'$ and integrating Eq.(\ref{72}), one
can obtain the evolutionary form as
\begin{equation} \label{73}
\varphi(a)-\varphi(0)=\int^{lna}_0\frac{1}{H}\sqrt{\frac{[27e^{2sx/3\nu}k\nu(\alpha-1)(\mu-\nu)
+X(s-3\alpha\nu)]N}{18\nu[X+9e^{2sx/3\nu}k(\nu-\mu)]}}dx.
\end{equation}

For the flat case $k=0$, assuming the initial condition
$\varphi(0)=0$, one can obtain
\begin{equation} \label{74}
\varphi(t)=\sqrt{\frac{(s-3\mu)(3\alpha\nu-s)}{18\omega\nu}}\frac{t^{3\alpha\nu/s}}{\alpha}~~~~~~(\alpha\neq0),
\end{equation}
\begin{equation} \label{75}
V(\varphi)=\frac{\nu(s-3\mu)(s-9\nu-3\alpha\nu)}{4\omega
s^2}\left[\sqrt{\frac{18\omega\nu}{(s-3\mu)(3\alpha\nu-s)}}\alpha\varphi\right]^
{2(1-s/3\alpha\nu)},
\end{equation}
The discussion about the physical statement of the potential is the
same as one in the new holographic quintessence model because the
both has the same potential.

In summary, by generalizing the previous work \cite{21,27} to the
NHDE model with $\rho_D=\frac{3\phi^2}{4\omega}(\mu H^2+\nu\dot{H})$
in the framework of Brans-Dicke cosmology, we have obtained the
evolution of EoS and given the present values of EoS $w_0$ (see
Tab.{\ref{T1}}),  which are basically consistent with the present
observation data $w_0=-1.10\pm0.14$. Furthermore, we have
established the correspondence between NHDE model with the
quintessence, tachyon, K-essence, dilaton, CG model in the non-flat
Brans-Dicke universe. Also, we have constructed the potentials and
the dynamics of these models, and found that the dynamics and
potential in the new holographic quintessence model are the same as
ones in the new holographic CG model, but they are completely
different from each other in the non-flat universe. For the new
holographic quintessence and CG models, if the parameters of the
potentials satisfy the constraints $3\alpha\nu/s=-1$,
$-1/2<\alpha<0$ and $\omega<-3/2$, the accelerated expansion can be
achieved in Brans-Dicke cosmology. It is worth stressing that not
only have we given some new results of the NHDE model in the
framework of Brans-Dicke theory, but also the previous results of
the new holographic dark energy in Einstein gravity \cite{20} can be
included as special cases of $\alpha=0$ or $k=0$ given by us, which
can describe the accelerated expansion.

\section*{Acknowledgments}

\small{Authors thank the anonymous reviewers for constructive
comments. This work has been Supported by the National Natural
Science Foundation of China (Grant No.10875056), the Natural Science
Foundation of Liaoning Province, China (Grant No.20102124), and the
Scientific Research Foundation of the Higher Education Institute of
Liaoning Province, China (Grant No.2009R35).}

\noindent{\footnotesize

\end{document}